# Baseline behaviour in human vision

**Authors:** Thomas Fabian[1]*

**Affiliations:**

[1]Department of History and Social Sciences, Technical University of Darmstadt; Darmstadt, Germany.

*Corresponding author. Email: thomas.fabian@tu-darmstadt.de

**Abstract:** Humans perceive their visual environment by directing their eyes towards relevant objects. The deployment of visual attention depends substantially on the stimulus's properties, higher cognitive processes, and biases and constraints of the visual system. Numerous models describe people's eye movements depending on the performed task or the viewed content. However, there is no universal, context-invariant model of human gaze behaviour. Here we show that statistical regularities can be utilised to model human gaze behaviour regardless of task, observer, and content. Using a context-agnostic eye movement model, we were able to describe human gaze behaviour better than a uniform random model in various viewing situations. Using a fixed transition kernel, the model can describe gaze patterns during reading, visual search, and scene perception, as well as for both adults and children. Thus, contrary to current belief, human gaze patterns follow a baseline behaviour, making them comparable across contexts. Since gaze behaviour is directly related to brain structure, our results provide the first evidence for the existence of an underlying, context-invariant motor prior in the human visual system.

**Keywords:** visual behavior, gaze dynamics, relative saccade lengths, computational modelling, cognitive neuroscience

# Introduction

Humans experience their visual surroundings through their eyes. To record visual information in detail, it must be located in the centre of the visual field. This foveal area accounts for only around one-thousandth of the human visual field[1]. The vast majority of the visual field is considered the periphery, where the density of photoreceptors decreases drastically[2] and the perceived resolution is even lower due to visual crowding[3]. For this reason, human visual sampling alternates between fixations, i.e., moments in which the eyes remain relatively still to absorb visual information, and eye movements to adjust the gaze, called saccades[4,5]. Humans perform saccades three to four times per second[6–8] to clearly see what seems most important at each point in time. Thus, human visual behaviour provides insights into the underlying decision-making processes that control attention.

Human visual behaviour is guided by three components: top-down processing, bottom-up processing, and the oculomotor system's characteristics[9]. Top-down processing is characterised by the task[5–7] and the viewer's attributes, e.g., experience or illness[10,11]. Bottom-up processing comprises the influence of low-level stimulus features, e.g., colours and patterns[12–15], and high-level features like meaning[8,16,17]. Top-down and bottom-up processing depend on the task, observer, and stimulus. In contrast, the oculomotor system's characteristics influence visual behaviour regardless of specific viewing circumstances. These characteristics include the human eye's structure, neurological processing constraints, and behavioural biases[18–21]. However, with only a few exceptions, the oculomotor system's influence cannot be used directly to model gaze behaviour. To date, centre-bias[18] and inhibition of return[19] are the only oculomotor-inspired components commonly applied in scanpath modelling. These components usually only have a supporting role as eye movements from different situations are considered not meaningfully comparable without taking into account top-down and bottom-up processes, i.e., the specific task and stimulus[8,22–24].

Consequently, research on modelling human scanpaths has been dominated by models focusing on single tasks and using all available stimulus information, as this yields the most accurate reproduction of human visual behaviour[25,26]. Machine learning approaches are widely used for modelling human scanpaths during scene perception[26,27] since they can incorporate various types of information without explaining how these precisely influence visual behaviour. Although many explicitly formulated models for scene perception include stimulus-independent biases[28,29], they also use saliency maps[12] to prioritise image areas based on low-level properties. Similarly, the modelling of gaze behaviour during reading often considers the text exclusively, with even stochastically motivated models requiring information such as word lengths and word frequencies[22,30,31]. There are scanpath models built purely on biases in gaze behaviour[32,33], which show that simple statistical regularities can describe a portion of human gaze behaviour. However, as these models are based on viewing biases in specific scene perception tasks, they are not applicable to other tasks, such as reading. Thus, existing models of human visual behaviour either require knowledge about stimulus features or are only applicable to single tasks.

This seemingly inevitable dependence of eye movement models on external information raises a fundamental question: Do eye-movement patterns follow any intrinsic, context-invariant regularity? If so, a baseline model of human gaze behaviour could capture inherent statistical structures of fixation sequences without recourse to any knowledge of the task, observer, or stimulus. Finding a universal baseline behaviour would constitute a theoretical probe into the core dynamics of the human visual system and establish a normative reference for context-dependent models. In this work, we advance exactly this perspective. Following the finding that the distribution of relative saccade lengths remains stable across tasks, observers, and stimuli[20], we propose and formalise a minimalist Markov chain eye movement model that rests solely on empirical regularities of eye movements relative to the preceding eye movement. Thus, our model constitutes a **B**aseline **U**sing **R**elative **R**esponses **I**ndependent of **T**ask, **O**bserver, and **S**timulus (BURRITOS). The BURRITOS model only tracks the last eye movement length and angle. Using this information, the model samples the next eye movement length from the distribution of relative saccade lengths[20]. It samples the angle from the statistical distribution of all eye movements, in which a memoryless inhibition-of-return[19] term suppresses only the direct return. By sampling eye movements from these empirical distributions, BURRITOS employs neither a saliency map, nor any centre-bias, nor does it involve memory. It operates as a zero-information baseline, free from assumptions about task, stimulus content, observer characteristics, or even the mere spatial layout.

To assess the generality and robustness of BURRITOS as a universal baseline, we assemble an unprecedented dataset comprising 36 eye-tracking experiments, in which 4,170 participants complete 233,584 experimental trials and perform 69,168,018 fixations. These experiments encompass a wide range of tasks, including reading, visual search, and scene perception, a diverse spectrum of stimulus types, sizes, and modalities, as well as various age groups and native languages. We focus on four experiment categories: reading, visual search, scene perception, and children. The reading experiments reflect different language families as they comprise Danish, Dutch, English, French, Estonian, Hebrew, and Russian. In the visual search experiments, participants find a specific shape among others, identify which object is not visible to another person, pick a described object, observe the appearance of a stimulus at the correct location, or fixate on an object upon bilingual instructions. In the scene perception experiments, participants look at faces and everyday objects, watch performed social scenes, view images while listening to spoken prompts, or view natural scenes, either without audio or accompanied by environmental sounds. To represent a clearly separated age group, we introduce an experiment category for children. Here, we focus on children under the age of five and mainly use data from the Peekbank repository[34]. Those comprise experiments on early word understanding and bilingualism. In addition to the data from Peekbank, we include experiments in which children watch performed social interactions or view images while listening to spoken prompts.

The contributions of this work are fivefold. First, we mathematically formulate a universal statistical regularity in human eye-movement behaviour, the distribution of relative saccade lengths[20]. Second, we introduce BURRITOS, a parsimonious, task- and stimulus-agnostic model for human scan path prediction that serves as a normative reference, establishing a foundation against which more specialised models can be compared. Third, to examine the broad applicability of our model, we compile the most extensive and heterogeneous corpus of eye-tracking data for evaluating eye movement models, which features a wide variety of tasks, observers, and stimuli. Fourth, we assess BURRITOS's validity by comparing its ability to predict empirical scanpaths to that of a model uniformly sampling eye movement lengths and angles. Fifth, by confirming the existence of a context-invariant baseline behaviour in human scanpaths, this work challenges the prevailing assumption that describing eye movement behaviour requires the consideration of task, observer, and stimulus. Instead, it suggests that a portion of gaze dynamics emerges from the visual system's intrinsic properties. Moreover, this finding would have direct neurobiological implications. It has been shown that fMRI activity can be reliably linked to eye-movement patterns[35], indicating that spatial priority maps are instantiated in the human brain. Given this link between brain activity and visual behaviour, finding a context-invariant component in human visual behaviour would also point to the existence of a default visual attention network hardwired in the human brain. This scaffold of spatial priority would generate consistent visual behaviour that can be modulated by task- and stimulus-driven signals atop a shared neural foundation.

# Results

The current work investigated to what extent an eye movement model based on relative saccade lengths can predict human gaze behaviour. In the following sections, we explain the structure of our BURRITOS model and describe how we evaluate its performance in comparison to a baseline model. After that, we describe how our model's performance is correlated with trial length and how its performance compares to that of the baseline model, both across all experiments in our eye-tracking corpus and for each experiment category separately.

## Model structure

In designing BURRITOS, we part ways with the assumptions of most eye movement models, which suggest that gaze behaviour primarily depends on low-level stimulus features and high-level cognitive processes. We estimate the recently found probability distribution of relative saccade lengths[20] to obtain a universal, context-agnostic, mathematically formulated relationship for the prediction of saccade lengths (see "Methods" for details). Based on this probability distribution, our Markov chain model performs saccades purely automatically, without any influence of the stimulus content, meaning, or the task completed by the viewer. The goal is to represent a universal, context-invariant motor prior. BURRITOS is not designed to represent visual behaviour accurately, but to investigate the existence of a universal motor prior in gaze behaviour and whether some eye movement properties can be explained independently of the viewing context through simple relational connections. The operating principle of BURRITOS is shown in Fig. 1 for a short saccade upwards to the right (Fig. 1a) and a longer saccade slanted downwards to the left (Fig. 1b).

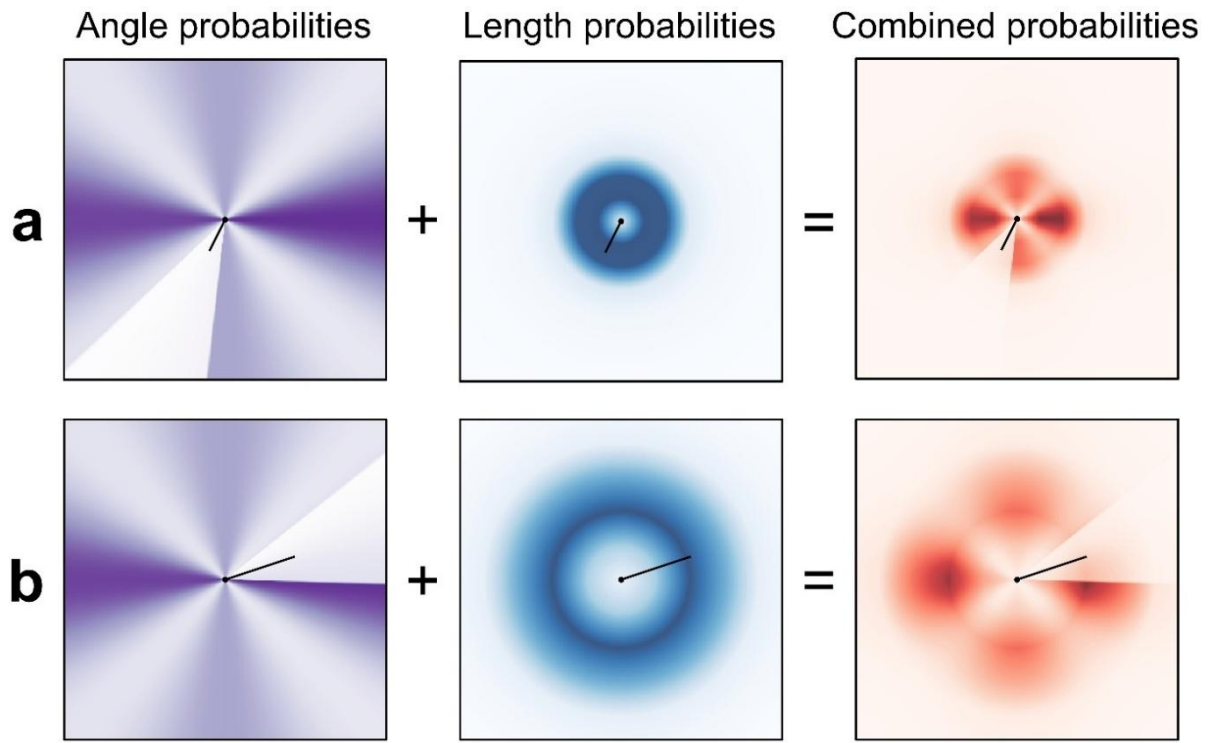

**Fig. 1** Probability allocation of BURRITOS. Examples of probability maps applied to previous eye movements that are **a** short and to the top-right, and **b** long, leftward, and slanted downward. The black line indicates the eye movement ending at the centre of a plot. More saturated colours represent higher probabilities. Angle probabilities are not limited by length but cut for visualisation. The angle probabilities follow the empirical distribution and apply an inhibition-of-return term, as indicated by the lighter area around the previous eye movement. The length probabilities follow the probability distribution of relative saccade lengths. The combined probabilities result from modulating angle and length probabilities.

Basing the eye movement length prediction of BURRITOS solely on the distribution of relative saccade lengths renders the transition probabilities to other states memoryless, which fulfils the Markov property and suggests an implementation via a Markov chain. This implementation involves viewing fixations as a random walk, where the length and angle of an eye movement depend solely on the previous eye movement. For the angle of a saccade, we use the empirical probability distribution of angles, which we modulate using an inhibition-of-return (Posner & Cohen) term depending on the previous angle (Fig. 1, left; see 'Methods' for details). The inhibition of return is memoryless in BURRITOS, only inhibiting the return in the direction of the last fixation. BURRITOS predicts the saccade length by sampling a relative saccade length from the estimated distribution and multiplying it by the previous saccade's length (Fig. 1, centre). For example, if a relative saccade length of two is sampled, BURRITOS predicts that the next saccade will be twice as long as the previous one. The combination of the probability distributions for angle and length yields the final spatial distribution, which BURRITOS attributes to an eye movement based on the previous saccade (Fig. 1, right). BURRITOS uses the human starting point of a trial as its only information, so it does not even consider the dimensions of a stimulus, as would be necessary if a centre-bias were used.

## Model evaluation

We evaluate the performance of BURRITOS using a second model that serves as a baseline. This model proceeds in a similar way to BURRITOS, but samples the saccade lengths and angles from uniform distributions. The probability of a saccade length is the same for all lengths from zero to the usual maximum of 20° or 600 pixels; for the angles, all angles are equally probable. We decide against the empirical distributions of the absolute saccade lengths, as these are highly task-dependent, and the baseline model should not have an information advantage over BURRITOS. We create this second model as a baseline because no existing model describes gaze behaviour across all task types in our data. We do not apply specialised models, as they naturally perform better in one task but require considerably more information and cannot be applied to other tasks.

To evaluate how well BURRITOS captures human eye movement patterns, we apply three metrics for a complementary assessment of discriminability, saliency alignment, and complete-distribution fidelity (see 'Methods' for details). With the Area Under the ROC Curve (AUC)[36], we measure the model's discriminative power in placing probability mass on empirical saccade lengths versus randomly sampled lengths. An AUC of 0.5 corresponds to chance-level performance, whereas values closer to one indicate that the model systematically ranks empirical saccade lengths higher than random ones. To evaluate whether BURRITOS assigns higher saliency to the exact locations where fixations occur, we employ the Normalised Scanpath Saliency (NSS)[37], which measures the alignment between the model's probability landscape and the empirical fixations. An NSS of zero implies that empirical fixations are on average positioned where the model predicts a mean probability. A positive NSS indicates that fixations tend to be located in regions that the model predicts to have high saliency. Lastly, we employ the Earth Mover's Distance (EMD)[38], or first-order Wasserstein distance, to quantify the full-distribution fidelity between predicted and empirical saccade length histograms. EMD is zero only if the two histograms are identical and grows with the distance over which the probability mass must travel to transform one histogram into the other. Thus, lower EMD values signify a smaller distance and a better alignment of the compared histograms. In our analyses, EMD represents the average number of pixels by which a predicted saccade differs from the empirical one, as we consider all eye-tracking data in pixels rather than visual angles.

To measure the performance of BURRITOS and the baseline model, we let both models predict the fixations for each empirical trial in our eye-tracking corpus. Thus, for each model, we obtain 233,584 scanpath predictions consisting of 84,771,195 fixations. Since we have empirical viewing data for each trial, we can determine the predictive quality of both models for each scanpath. To evaluate performance across all experiment categories, we calculate the mean values per dataset, yielding 36 data points per model. For the evaluation by category, we average the data per participant, as statistical robustness would not be given with only eight datapoints. We report the p-values adjusted with the false-discovery-rate correction[39] at the conventional significance level ($\alpha = 0.05$), together with Cohen's d[40] to assess the effect size.

# Correlation of model performance with trial length

We suspect that BURRITOS's prediction performance for short eye movement sequences is close to the chance level since it is a purely stochastic model. Therefore, we analyse the correlation of the performance metrics with the average length of a predicted trial. The results of the correlation analysis are shown in Fig. 2.

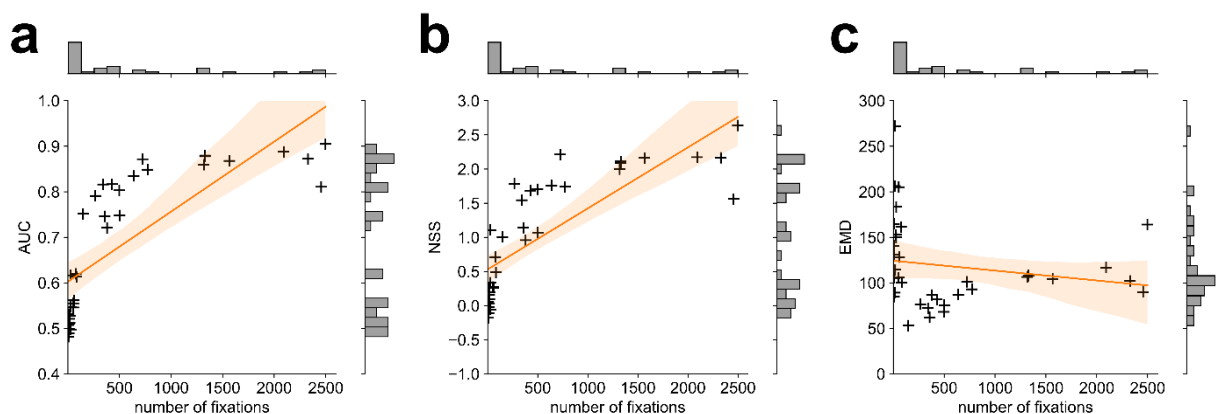

**Fig. 2** Correlation of BURRITOS's performance with the predicted trial's length. The performance is measured by **a** Area Under the ROC Curve (AUC), **b** Normalised Scanpath Saliency (NSS), and **c** Earth Mover's Distance (EMD). The x-axis shows trial length measured by the number of fixations, and the y-axis shows the score achieved by BURRITOS. Each cross represents the score for scanpath predictions, averaged across all participants and trials, for each experiment in the eye-tracking corpus (n = 36). For AUC and EMD, higher values indicate better performance of BURRITOS. EMD decreases with better performance. The orange line represents the ordinary-least-squares regression of the correlation, and the shaded area represents its 95% confidence interval.

We find that AUC increases strongly with trial length (Spearman's $\rho = 0.94$, $p < 0.001$; Fig. 2a), indicating that BURRITOS can better differentiate between empirical lengths and random distractor lengths. For short trials, the performance is close to the chance level of 0.5, indicating that BURRITOS provides no information for these trials. When predicting trials containing more than 100 fixations, our model achieves AUC values of at least 0.7, so its predictions' precision is considerably above the chance level. NSS likewise exhibits a robust positive correlation with trial length (Spearman's $\rho = 0.91$, $p < 0.001$; Fig. 2b), showing that empirical saccades fall further above the model's baseline expectation as more saccades are predicted, signifying tighter alignment between the model's high-confidence predictions and actual behaviour. Similar to the AUC, the results for the NSS show that in short trials, the positions of empirical saccades only land on locations ascribed mean saliency by the model. For longer trials, BURRITOS assigns the regions in which empirical saccades occur average saliency values that are 1 to 2.5 standard deviations above the mean.

By contrast, EMD displayed a moderate, non-significant negative correlation with trial length (Spearman's $\rho$ = -0.32; $p$ = 0.057; Fig. 2c). This suggests that longer trials do not or only slightly reduce the misallocated mass between predicted and empirical histograms, although the effect is relatively weak and not significant. The length-invariant EMD implies that any mismatch in the shape of the distributions is difficult to offset with longer trials. The EMD is lowest with values between 50 and 80 when a trial is 100 to 500 fixations long. For longer trials, the EMD values even increase again, resulting in an average saccade deviating more than 100 pixels from the length of the empirical saccade.

Overall, the results of the correlation analysis show that BURRITOS can predict the scanpaths in trials with few fixations only with chance-level performance. However, if the trials are longer, the scanpaths predicted by BURRITOS converge towards human behaviour, at least for ranking lengths and saliency prediction. We see this in the increase and saturation of the values achieved for AUC and NSS. However, the overall distribution fidelity of the predicted scanpaths, as measured by the EMD, does not decrease. In practice, this means that increasing the trial length primarily improves BURRITOS's ability to predict ranking and saliency. In contrast, its ability to increase geometric similarity and capture finer distributional nuances remains essentially unchanged.

## Baseline comparison across all experiments

For the evaluation of BURRITOS's performance in predicting human scanpaths, we first consider all experiments. Figure 2 shows comparisons of BURRITOS's performance against the uniform baseline across the three metrics AUC, NSS, and EMD. The data show that for AUC (Fig. 3a) and NSS (Fig. 3b), the dataset-by-dataset covariance remains below unity, indicating that the baseline model cannot match BURRITOS's performance for these measures. Correspondingly, the covariance of the EMD scores (Fig. 3c) tilts downward, signifying that datasets with a lower, thus better, BURRITOS EMD correspond to a higher baseline EMD. The plots show that our model not only exceeds the baseline in ranking true saccades and aligning them with its saliency predictions but also produces closer overall distributions.

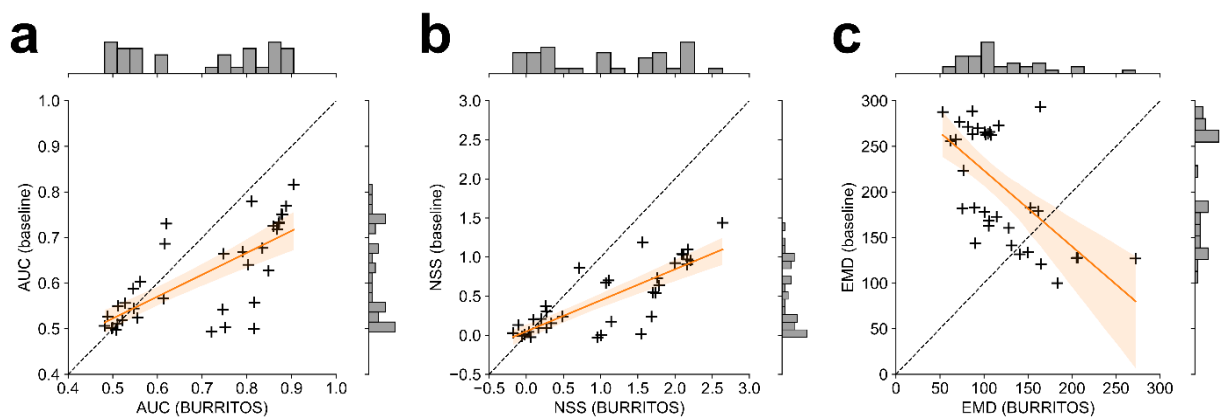

**Fig. 3** Comparison of BURRITOS's performance with the uniform baseline model. Performance is measured by **a** Area Under the ROC Curve (AUC), **b** Normalised Scanpath Saliency (NSS), and **c** Earth Mover's Distance (EMD). The x-axis indicates the score achieved by BURRITOS, the y-axis the score of the baseline model. Each cross represents the scores for scanpath predictions, averaged across all participants and trials, for each experiment in the eye-tracking corpus (n = 36). For AUC and EMD, crosses located below the dashed diagonal line of equality indicate better performance of BURRITOS. For EMD, crosses above the diagonal imply better performance, as EMD decreases with better performance. The orange line represents the ordinary-least-squares regression of both models' scores, and the shaded area represents its 95% confidence interval.

Across datasets, we test whether BURRITOS achieves significantly better AUC, NSS, and EMD scores than the uniform baseline. For AUC, BURRITOS significantly outperforms the baseline model (Welch's $t = 2.51$, $p = 0.007$, $d = 0.60$), indicating that it discriminates empirical saccades from distractors significantly better than the uniform baseline model. Similarly, NSS is significantly greater for BURRITOS's predictions (Welch's t = 3.58, p < 0.001, d = 0.86), reflecting that empirical fixations align with the model's saliency peaks more tightly than with the saliency peaks of the baseline model. Thus, BURRITOS appears to better concentrate probability around empirical lengths rather than spreading it uniformly. Additionally, EMD is significantly lower, i.e., better, for BURRITOS compared to the baseline (Welch's t = -6.69, p < 0.001, d = 1.60). This confirms that our model achieves a substantially closer match between the predicted and empirical saccade-length histograms across all experiments.

These results statistically confirm that our model not only ranks and localises human saccade lengths more accurately than the uniform baseline model, but also achieves better distributional fidelity. Hence, the baseline comparison across all experiments shows that BURRITOS, given large enough trials for its behaviour to converge, overcomes the initial randomness with its relative saccade length transition kernel and systematically improves, sharpens, and spatially aligns its predictions with human eye movement patterns.

## Baseline comparison by experiment category

Across the four experiment categories, reading, visual search, scene perception, and children, we compare our BURRITOS's performance with the uniform baseline model on the three metrics AUC, NSS, and EMD. The comparison of the two models broken down by category is carried out at the participant level to ensure statistically robust results. The Wilcoxon signed-rank test is used for all model comparisons within each experiment category.

The AUC values of BURRITOS are significantly higher than those of the baseline model, for reading (W = 290,728, p < 0.001, d = 0.30), visual search (W = 121,220, p < 0.001, d = 0.31), scene perception (W = 140,784, p < 0.001, d = 0.78), and children (W = 1,969,858, p < 0.001,

d = 1.75). BURRITOS also achieves significantly higher NSS scores than the baseline model for reading (W = 341,650, p < 0.001, d = 0.50), visual search (W = 141,243, p < 0.001, d = 0.46), scene perception (W = 142,823, p < 0.001, d = 0.86), and children (W = 1,978,261, p < 0.001, d = 2.41).

In contrast, as shown in Fig. 4, BURRITOS only achieves significantly lower EMD scores for experiments in the categories reading (W = 15,924, p < 0.001, d = 1.39), scene perception (W = 7,537, p < 0.001, d = 1.29), and children (W = 39,668, p < 0.001, d = 2.34). For viewing behaviour during visual search, the distributional fidelity of BURRITOS's predictions is not significantly better than the uniform baseline model's (W = 113,820, p = 0.96, d = –0.07).

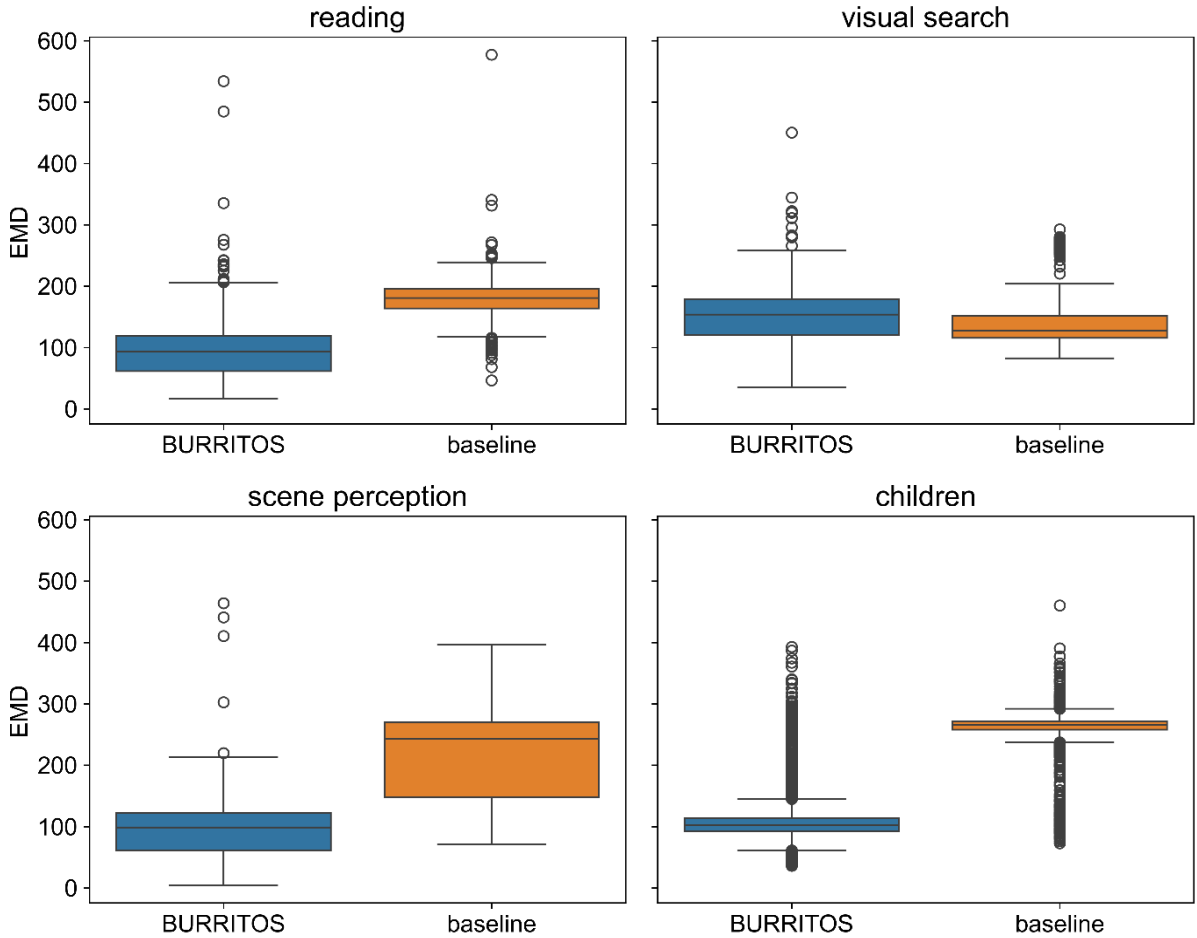

**Fig. 4** Category-wise comparison of BURRITOS's EMD scores with those of the uniform baseline model. Each box plot shows the performance of BURRITOS (blue) and the baseline model (orange) for one of the four experiment categories: reading, visual search, scene perception, and children. The y-axis shows the EMD values, which decrease with better performance. The values show the mean EMD score over all trials of a single participant. Each outlier, indicated by a circle beyond the box plots' whiskers, represents the mean EMD score for a single participant.

Together, these results demonstrate that BURRITOS consistently outperforms a uniform baseline in ranking, saliency alignment, and overall distributional fidelity across diverse experiments involving reading, scene perception and children. For visual search, the model does not narrow or reshape the full-length distribution, but it captures the discriminative and saliency aspects of human eye movements.

# Discussion

In this work, we demonstrated that human visual behaviour can be explained across tasks, observers, and stimuli with a simple regularity in gaze dynamics. We presented BURRITOS, the first model capable of predicting human eye movement patterns across a wide range of tasks and stimuli. Thus, this information-agnostic, parsimonious eye movement model constitutes the first universal baseline for human gaze allocation. Unlike existing models that incorporate knowledge of task demands, observer characteristics, stimulus properties, or even implicit priors such as centre-bias, BURRITOS operates with no contextual information. To predict eye movements, our model utilises a fixed and explicitly formulated transition kernel that applies to any dataset without requiring retraining, adaptation, or tuning. Despite its structural simplicity, BURRITOS consistently outperforms a uniform random model on various evaluation metrics. When analysed for experimental domains individually, BURRITOS consistently outperforms the baseline at predicting visual behaviour for reading, scene perception, and children. For visual search, our model outperformed the uniform baseline in AUC and NSS, with only the EMD showing no improvement.

First, we mathematically formalised a recently discovered context-invariant regularity in human gaze behaviour, the distribution of relative saccade lengths[20], as a transition kernel. This kernel provides a mathematically explicit, universal descriptor of a scanpath regularity, which we employed to create BURRITOS. To validate the model's generality, we assembled an eye-tracking corpus of unprecedented size and heterogeneity, comprising 36 independent experiments, over 4,000 participants, more than 230,000 trials, and over 69 million fixations. This expansive scope encompasses a broad range of viewing behaviours, including static scene viewing, face recognition, visual search, reading, and video viewing. Thus, it provides a general account of viewing behaviour, which makes it suitable for validating BURRITOS's performance and zero-shot generalisation. Without retraining, BURRITOS outperforms the uniform baseline model across all experimental domains, except for the EMD during visual search, which matches.

Besides its predictive power, the information-agnostic BURRITOS provides explanatory value given its parsimonious structure. Conceptually, the performance of BURRITOS across a wide range of tasks, observers, and stimuli invites a reconsideration of how the human visual system is organised. Finding a single kernel of saccade transitions that provides predictive information, regardless of context, supports the idea that context-invariant oculomotor structures underpin human gaze behaviour. Hence, our results contradict the prevailing assumption that low-level stimulus features and high-level cognitive processes drive visual behaviour, with oculomotor components only being a supplement. Instead, a universal oculomotor prior appears to constitute baseline visual behaviour since a substantial fraction of scanpath structure emerges from the visual system's intrinsic structure. As this baseline behaviour applies regardless of task- and stimulus-specific cognitive factors, these factors seem only to determine the specifics in which intrinsic motor tendencies manifest themself. Thus, top-down and bottom-up processes appear not to instantiate new transition rules but rather influence eye movement control by modulating a shared, low-dimensional baseline behaviour.

This finding also has a direct neurobiological implication. Given that research has used fMRI activity to predict eye-movement patterns[35], our finding indicates that spatial priority maps are instantiated in the human brain. The universal saccadic regularity captured by BURRITOS implies the existence of a default visual attention network, a hardwired circuit yielding a consistent behavioural foundation regardless of task or stimulus. Such an underlying network would provide a stable scaffold of spatial priority, with context-specific influences merely modulating the baseline behaviour. Investigating this network with neuroimaging could reveal the neurological and functional core of active human vision, shedding light on how brain regions interact to produce the baseline visual behaviour we observe.

The universal, context-free nature of BURRITOS opens promising avenues, especially in clinical applications and computer vision. In healthcare, deviations from the normative transition kernel of relative saccade lengths could serve as sensitive biomarkers for neurodegenerative diseases or forms of neurodiversity, like Alzheimer's disease or autism spectrum disorder, where subtle alterations in oculomotor control often precede overt cognitive symptoms[41,42]. By tracking patients' eyes when performing simple tasks, their relative saccade length distributions can be compared against the normative distribution to detect early signs of pathology with a low-resource procedure. In computer vision, integrating BURRITOS's transition kernel as a default saliency prediction module could improve algorithms in applications as diverse as autonomous driving, automated satellite data analysis, or medical imaging without the need for any domain-specific training data. Given its universal, lightweight, and parameter-free nature, BURRITOS is suitable for zero-shot deployment even in settings where domain-specific training data may be limited or unavailable.

For research on human visual cognition and gaze prediction models, BURRITOS constitutes a benchmark that defines the minimal, information-agnostic performance level any gaze model must exceed to claim an advance in explanatory power. Improvements beyond our model can be attributed to content-driven, cognitive, or task-specific mechanisms. BURRITOS itself could be extended to account for other universal cognitive mechanisms. BURRITOS currently considers only first-order spatial transitions. Extensions could incorporate fixation durations or memory components, improving the model's performance and conceptual scope. Furthermore, integrating BURRITOS with image-computable saliency would yield a hybrid model that combines universal dynamics with content sensitivity. Such an integration could considerably improve gaze predictions while retaining zero-shot generalisation.

In sum, we mathematically formulated a universal regularity in gaze patterns, introduced BURRITOS as a context-agnostic eye movement model, compiled an unprecedentedly large and heterogeneous eye-tracking corpus for scanpath model validation, and demonstrated BURRITOS's performance across four major viewing paradigms. Thereby, this work establishes a new foundation for gaze prediction research with substantial implications for visual cognition, neuroscience, clinical settings, and computer vision. BURRITOS captures a fundamental motor regularity underpinning visual behaviour in various contexts, providing a universal benchmark for models of attention and vision. As such, it offers a new angle for investigating the interplay between oculomotor constraints and cognitive control in human visual behaviour by demonstrating the existence of baseline behaviour in human gaze patterns.

## Methods

### Data

When creating the eye-tracking corpus, we follow the procedure that was used to investigate the regularity of relative saccade lengths[20]. Thus, we also use the four experiment categories: reading, visual search, scene perception, and children. With the three different task types, we represent different viewing situations, and with the children, we cover a distinctly different age group. The data from all 36 experiments analysed in this work are publicly available[34,35,43–72] (note that some publications report on more than one experiment). Each experiment complies with the respective ethical regulations, and every participant or a legal guardian provided written informed consent. To construct the scanpaths, each experiment's dataset either provides the raw eye-tracking data or chronological lists of all participants' fixations. We use the software EyeLink DataViewer (SR Research Ltd.) to transform raw eye-tracking data into chronological fixation lists. Additionally, for the empirical scanpaths, we neither use preprocessed data nor edit the data in any way.

The eye-tracking data comprises 4170 participants, with 2176 being adults and 1994 children. For the adults, data were recorded from 951 participants during reading, 660 during visual search, and 565 during scene perception. Each of these three task types comprises eight different experiments. For children, we include twelve experiments, as these experiments are highly inhomogeneous for two reasons. First, the cognitive abilities are considerably different within the age range, e.g., six-month-olds' cognitive skills are substantially less developed than three-year-olds'. Second, experiments with children commonly employ a mix of visual search, scene perception, and reading of single words. As many eye-tracking datasets do not mark which task was performed at any given time, we cannot differentiate children's gaze behaviour by task. Additionally, more data is necessary for children, as many experiments are shorter than those for adults because children tend to lose interest in experiments more quickly and struggle to maintain their attention.

Some biases in the eye-tracking corpus arise from the availability of suitable and publicly accessible data. First, most data comes from participants sitting down and viewing still images. No participants walked around during the experiment, and only in two cases did they view videos[63,71]. Second, certain people and cultures are underrepresented, as we were unable to access suitable data for experiments from regions such as Africa, Asia, Australia, or Latin America. Third, of the compared languages, only two out of seven are non-Latin (Russian and Hebrew), and Hebrew is the only one that is not read from left to right. Fourth, the children's data is biased towards US-American participants, with ten of the twelve experiments conducted in the US. Only one experiment was conducted in Canada with French-English bilingual children[58] and another one in Germany with German native speakers[72]. We can better counterbalance the dominance of US-American data in visual search and scene perception, with only 4 of 8 and 3 of 8 experiments from the USA, respectively.

## Distribution estimation

For estimating the distribution of relative saccade lengths, we use the data and preprocessing procedure from the original publication on relative saccade lengths[20]. For each trial of each participant in a dataset, we use the temporally sorted fixation positions to determine the absolute length of each eye movement. We divide every eye movement length by the length of the preceding eye movement, yielding n-1 relative saccade lengths for a trial with n eye movements. To ensure that the influence of all datasets on the results is equal and does not depend on their size, we determine the mean values for each dataset separately. For relative lengths in the spectrum from 0 to 20, we determine the frequency of occurrence for 500 discretised, equidistant values in each dataset. We normalise these frequencies with the number of all datapoints in the corresponding dataset. Using all datasets, we calculate the mean value of the relative frequency for each of the 500 points.

In the original examination of the distribution of relative saccade lengths, the focus was solely on the relative saccade lengths greater than one, which follow a power-law[20]. We estimate the entire distribution in two parts. Similar to the original study, we check which of the functions, lognormal, gamma, exponential, and power-law, best describe the part from 0 to 1 and which best describe the part from 1 to 20. For a quantitative assessment, we calculate the log-likelihood, the Akaike Information Criterion[73], and the Bayesian Information Criterion[74]. The best-fitting functions are the exponential function for the first part of the distribution from 0 to 1, and the power-law for the part from 1 to 20. We fit the parameters of the exponential function by linear regression using the Least Squares method[75,76]. For estimating the parameters of the power-law, we employ the Levenberg-Marquardt algorithm[77,78] since the Least Squares method does not converge. For the probability of relative saccade lengths smaller than one, we obtain the exponential function

$$y = 0.023 * \exp(-1.13 * (x-0.59)^2),$$

and for relative saccade lengths greater than one, the power-law function

$$y = (x*6.73)^{-2.03}.$$

With this estimation, we obtain a complete probability distribution for the relative saccade lengths, which we can use to stochastically sample the length of a saccade based on the length of the previous saccade. We perform the distribution estimation in Python.

## Model structure

BURRITOS is a Markov chain model whose state is defined only by the position of the current fixation and the length and angle of the last eye movement. For each trial to be predicted, we initialise this Markov chain with the first fixation of the human observer, one pixel for the length of the last eye movement, and zero degrees, i.e. a movement to the right, for the last angle. We use the length of one pixel because the average eye movement length strongly depends on the viewer and the viewing situation. We choose the shortest possible eye movement, as short saccades appear in most scanpaths. Initialising BURRITOS with the value zero would, in combination with the prediction via the relative saccade lengths, lead to constant prediction of saccade length zero.

Our model predicts the saccade length based on the length of the previous saccade. BURRITOS samples a value from the probability distribution of the relative saccade lengths and multiplies it by the length of the previous saccade. For the angle, we use a von Mises distribution that reflects the empirical distribution of the absolute angles of all eye movements. To capture the dominance of horizontal and vertical eye movements, we place peaks at

$$\mu_f = 0,\ \mu_o^+ = +\pi/2,\ \mu_o^- = -\pi/2,\ \mu_r = \pi,$$

so that the base density is

$$g(\theta) = w_f\, f_{vM}(\theta;\ \mu_f,\ \kappa_f) + w_o[f_{vM}(\theta;\ \mu_o^+,\ \kappa_o) + f_{vM}(\theta;\ \mu_o^-,\ \kappa_o)] + w_r\, f_{vM}(\theta;\ \mu_r,\ \kappa_r),$$

where each von Mises PDF is

$$f_{vM}(\theta;\ \mu,\ \kappa) = 1/[2\pi\, I_0(\kappa)] * \exp[\kappa \cos(\theta - \mu)].$$

For the angles, we decide against using relative angles, as these struggle to reproduce the apparent dominance of horizontal and vertical eye movements. However, BURRITOS also includes the last angle when predicting the next eye movement angle, as the inhibition-of-return term depends on the angle of the last eye movement. The implementation of this term ensures that the probabilities of the angles that would lead to the last fixation point are reduced. Specifically, the probabilities from the distribution of the absolute angles in a range of 40° around the direction to the last fixation are reduced by 70%. For example, if the last eye movement had an angle of 90°, i.e. went straight downwards, the probabilities for an upward movement would be less likely as the probabilities for the angles from 250° to 290° are reduced by 70%. After applying the inhibition-of-return term, we renormalise so the total probability distribution integrates to one.

The size of the return range and the inhibition strength are arbitrary and can, in principle, be chosen differently. We define the range of 40° opposite to the last angle as the return, as we have to suppress the return with a uniform angle for any saccade length. A larger angle would be appropriate for particularly short saccades, as the viewed object is closer and occupies a larger angle. Similarly, for long saccades, a smaller angle would better describe the return range. Similarly, we select an inhibition strength of 70%, as returns in the 40° range must still be possible, for example, when skipping to the beginning of the next line while reading. BURRITOS applies the same mechanisms in all situations, regardless of task, observer, and stimulus.

## Performance evaluation

To evaluate the model's performance, we partition the saccade-length range into M discrete equal-length bins, which we use to form the histograms for the model's predictions and the empirical saccade lengths. Then $p_j$ denotes the model's predicted probability for a saccade to be in the j-th bin, $p'_j$ is the empirical probability in that same bin, and $b_i \in \{1,…,M\}$ is the bin index containing the i-th empirical saccade. In our analysis, we set M=30 to balance sufficient resolution to capture the distribution's shape and an over-fragmented saccade length interval.

To assess how well the predicted saccade length distributions and the empirical distributions align, we compute three measures: Area Under the ROC Curve (AUC), Normalised Scanpath Saliency (NSS) and Earth Mover's Distance (EMD). For each trial of N empirical saccades, we record the bin index $b_i$ of each observed length. To obtain AUC, we regard the probability our model assigns to the bin containing the i-th empirical saccade $p_{b_i}$ as a positive score and sample a negative score by uniformly drawing an index from {1…M}. The AUC denotes the probability that a randomly chosen positive score exceeds a randomly chosen negative score. As this classification is binary, an AUC of 0.5 corresponds to chance and 1 indicates perfect discrimination.

To compute NSS, we normalise the model distribution with the z-score

$$z_j = (p_j - p_{mean}) / \sigma_p,$$

where $p_{mean}$ and $\sigma_p$ are its mean and standard deviation across all bins. NSS denotes the mean $z_{b_i}$ over all empirical saccades, with positive NSS indicating that observed saccades fall in bins above the model's mean probability. NSS measures the difference between the estimated probability and the mean probability in units of $\sigma$).

The EMD (equivalent to the first-order Wasserstein distance) measures the work required to transform the predicted saccade length distribution into the empirical distribution along the length axis as

$$EMD = \sum_{j=1…M} |p_j - p'_j| * \Delta c,$$

where $\Delta c$ is the bin width. We multiply the summed distances by $\Delta c$ to transform the number of bins back into pixels, which we use to measure saccade lengths. EMD is zero only when the predicted and empirical lengths are equal, and increases with the amount of misplaced probability mass and the distance it must move. To test for significant differences between the scores of BURRITOS and the uniform baseline model on all three measures, we apply the Shapiro-Wilk test and Levene's test to determine the appropriate statistical test. We create the plots with matplotlib and seaborn in Python.

# Data availability

All datasets analysed in this work are publicly available in the repositories of the respective publications.